\documentclass[conference]{IEEEtran}

\usepackage[cmex10]{amsmath}
\usepackage{amssymb}

\usepackage{cases}
\usepackage{multirow}
\usepackage{empheq}

\usepackage{cite}
\usepackage{verbatim}

\ifCLASSINFOpdf
  \usepackage[pdftex]{graphicx}
\else
  \usepackage[dvips]{graphicx}
\fi
\usepackage[caption=false,font=footnotesize]{subfig}

\usepackage{color}
\usepackage{multirow}

\newtheorem{theorem}{Theorem}
\newtheorem{lemma}[theorem]{Lemma}

\newtheorem{corollary}[theorem]{Corollary}
\newtheorem{definition}[theorem]{Definition}

\newtheorem{remark}[theorem]{Remark}

\begin{document}

\sloppy

\title{Locally Rewritable Codes for Resistive Memories}

%\author{
%\IEEEauthorblockN{Yongjune Kim and B. V. K. Vijaya Kumar}
%\IEEEauthorblockA{Data Storage Systems Center (DSSC)\\
%    Carnegie Mellon University\\
%    Pittsburgh, PA, USA\\
%    Email: yongjunekim@cmu.edu, kumar@ece.cmu.edu}
%\and
%\IEEEauthorblockN{Robert Mateescu}
%\IEEEauthorblockA{HGST Research\\
%San Jose, CA, USA\\
%Email: robert.mateescu@hgst.com}
%}

\author{\IEEEauthorblockN{Yongjune Kim\IEEEauthorrefmark{1},
	Abhishek A. Sharma\IEEEauthorrefmark{1},		
	Robert Mateescu\IEEEauthorrefmark{2},
	Seung-Hwan Song\IEEEauthorrefmark{2},
	Zvonimir Z. Bandic\IEEEauthorrefmark{2},\\
	James A. Bain\IEEEauthorrefmark{1},
	and
	B. V. K. Vijaya Kumar\IEEEauthorrefmark{1}}
\IEEEauthorblockA{\IEEEauthorrefmark{1}Data Storage Systems Center (DSSC), Carnegie Mellon University, Pittsburgh, PA, USA\\ Email: \{yongjunekim, abhisheksharma\}@cmu.edu, \{jbain, kumar\}@ece.cmu.edu}
\IEEEauthorblockA{\IEEEauthorrefmark{2}HGST Research, San Jose, CA, USA\\
Email: \{robert.mateescu, seung-hwan.song, zvonimir.bandic\}@hgst.com}
}

%% To balance the two columns, you should reduce the text-height of
%% the last page using the following command:
%%%%%%%%%%%%%%%%%%%%%%%%%%%%%%%%%%%%%%%%%%%%%%%%%%%%%%%%%%%%%%%%%%%%%
%\addtolength{\textheight}{-9.35cm}
%%%%%%%%%%%%%%%%%%%%%%%%%%%%%%%%%%%%%%%%%%%%%%%%%%%%%%%%%%%%%%%%%%%%%
%% with an appropriate value. This command must be place on the second
%% last page, i.e., for a one-page abstract here, for a two-page
%% abstract right after the \maketitle command.

%% Create the title:
\maketitle

\begin{abstract}
	We propose locally rewritable codes (LWC) for resistive memories inspired by locally repairable codes (LRC) for distributed storage systems. Small values of \emph{repair locality} of LRC enable fast repair of a single failed node since the lost data in the failed node can be recovered by accessing only a small fraction of other nodes. By using \emph{rewriting locality}, LWC can improve endurance limit and power consumption which are major challenges for resistive memories. We point out the duality between LRC and LWC, which indicates that existing construction methods of LRC can be applied to construct LWC. 

\end{abstract}

\section{Introduction}
	In the big data era, coding for storage systems has become more important than before. Recently, coding for distributed storage systems has become an attractive research area at the (higher) system level. In addition, coding for nonvolatile memories and hard disk drives (HDD) is also important to achieve high-density storage systems at the (lower) physical level. 
	
	An important group of codes for distributed storage system is locally repairable (or recoverable) codes (LRC)~\cite{Huang2007pyramid, Gopalan2012}. An $(n, k, d, r)$ LRC is a code of length $n$ with information (message) length $k$, minimum distance $d$, and repair locality $r$. If a symbol in the LRC-coded data is lost due to a node failure, its value can be repaired (i.e. reconstructed) by accessing at most $r$ other symbols~\cite{Gopalan2012, Tamo2014LRC}. 
	
	One way to ensure fast repair is to use low repair locality such that $r \ll k$ at the cost of minimum distance $d$. The relation between $d$ and $r$ is given by~\cite{Gopalan2012}
	\begin{equation} \label{eq:general_singleton}
		d \le n - k - \left\lceil \frac{k}{r} \right\rceil + 2.
	\end{equation}
	It is worth mentioning that this bound is a generalization of the Singleton bound. The LRC achieving this bound with equality are called \emph{optimal}. Constructions of the optimal LRC were proposed in~\cite{Silberstein2013, Tamo2014LRC, Tamo2013matroid}. Recently, several binary LRC constructions have been proposed~\cite{Goparaju2014, Shahabinejad2014, Huang2015, Tamo2015subcodes, Silberstein2015}. 
	
	At the lower (physical) level, coding for nonvolatile memories is an active research area since nonvolatile memories including flash memories and resistive memories are important parts of mobile devices and solid state drives (SSD). 
	
	In this paper, we investigate coding for resistive memories including phase change memories (PCM) and resistive random-access memories (RRAM). Resistive memory technologies are promising since they are expected to offer higher density than dynamic random-access memories (DRAM) and better speed performance than NAND flash memories~\cite{xpoint2015}. 
	
	The major challenges of resistive memories are endurance limit and power consumption~\cite{Wong2010PCM, Wong2012RRAM}. Endurance limit refers to the maximum number of writings that the memory can endure. In order to improve endurance and power consumption of such memories, we propose locally rewritable codes (LWC).\footnote{LWC instead of LRC is used as the acronym of locally rewritable codes in order to distinguish them from locally repairable codes (LRC).} Inspired by the \emph{repair locality} defined for distributed storage systems, we introduce the \emph{rewriting locality} which improves power consumption and endurance limit. In addition, we show the \emph{duality} between LRC and LWC, which indicates that existing construction methods of LRC can be used to construct LWC. 

	The rest of this paper is organized as follows. Section~\ref{sec:background} explains the basics and challenges of resistive memories. Section~\ref{sec:channelmodel} presents the notation and the defect channel model for resistive memories. In Section~\ref{sec:lwc}, we propose LWC and explain the duality of LRC and LWC. In Section~\ref{sec:conclusion}, we will discuss the future work and conclude the paper. 

\section{Resistive Memories}\label{sec:background}

PCM and RRAM are two major types of resistive memories. Both have attracted significant research interest due to their scalability, compactness, and simplicity. The main challenges that prevent their large-scale deployment are endurance limit and power consumption~\cite{Wong2010PCM, Wong2012RRAM}. The endurance limit refers to the maximum number of writes before the memory becomes unreliable. As explained in following subsections, the resistive memory cells have the limited endurance. Beyond this number, these cells can become stuck-at defects. In addition, the power consumption depends on the number of writes. 

\subsection{Phase Change Memories (PCM)}

\begin{figure}[!t]
	\centering
	\includegraphics[width=0.5\textwidth]{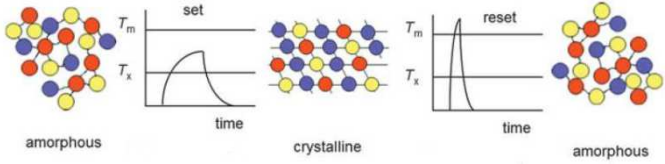}
	\caption{Principle of PCM. Starting from the amorphous phase with large resistance, a current pulse is applied. After sufficiently long pulse heats the material above the minimum crystallization temperature $T_{\text{x}}$ to crystallize the material, the resistance is low (SET operation). After the larger and short pulse is applied to heat the material above the melting temperature $T_{\text{m}}$, the material is melt-quenched and returns to the amorphous (RESET operation)~\cite{Raoux2014PCM}.}
	\label{fig:PCM}
	\vspace{-3mm}
\end{figure}

PCM consists of chalcogenide materials like Ge-Sb-Te (GST), which are known to have two stable resistance states~\cite{Wong2010PCM}. As shown in Fig.~\ref{fig:PCM}, the low resistance state (LRS) corresponds to a crystalline structure of the chalcogenide material, whereas the high resistance state (HRS) corresponds to an amorphous structure. The transition from HRS to LRS, known as SET, is brought about by applying a long and low-power heat pulse to the device by the means of a heating element. Similarly, the transition from LRS to HRS, or RESET, is brought about by pulsing the device with a short and high-power heat pulse that melts the chalcogenide, thus amorphizing it. Both operations can be done on the nanosecond time scale. However, the elapsed time for SET operation could be up to ten times of RESET operation~\cite{Wong2010PCM, Raoux2014PCM}.

PCM has shown great promise as a storage-class memory due to its superior resistance ratio, scalability, low-energy switching, and high-speed ~\cite{Raoux2008, Raoux2014PCM}. However, one of the main challenges for PCM is its endurance limit. From the point of view of the data, this corresponds to stuck-at defects (or stuck-at faults). Such defects may either appear in as-fabricated devices due to process variations or may be generated during the cycling process, i.e., rewriting.

The stuck-at defects in PCM are classified into: (1) stuck-at LRS defect which corresponds to the device in LRS being unable to RESET to HRS; and (2) stuck-at HRS defect which corresponds to the device in HRS incapable of being SET to LRS for the same operating conditions~\cite{Kim2012PCM}. The stuck-at LRS defect is traditionally attributed to the formation of crystallites in the amorphous state that do not melt (during the amorphization pulse) due to local inhomogeneities~\cite{Lee2014PCM}. This causes the HRS to gradually move towards the LRS with cycling. Similarly, the stuck-at HRS is attributed to the formation of voids in the materials and their eventual agglomeration~\cite{Chen2009PCM}. This causes the material to experience an inhomogeneous and often insufficient heating during the SET operation. 

%In the past, some methods proposed to mitigate these failures have involved limiting the peak programming current~\cite{Kim2012PCM} or corrective measures like increasing the operating power of the cell. However, these methods do not serve as permanent solutions to making entire system reach acceptable endurance levels. This is due to the inherent stochastic nature of the process of formation of crystallites (stuck-at LRS defects) and voids (stuck-at HRS defects).

\subsection{Resistive Random-Access Memories (RRAM)}

\begin{figure}[!t]
	\centering
	\includegraphics[width=0.45\textwidth]{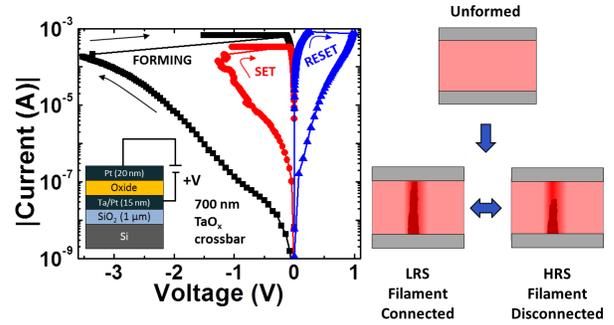}
	\caption{Direct current--voltage characteristics of the RRAM device showing forming and switching processes and a physical mechanism of filament formation and dissolution.}
	\label{fig:RRAM}
	\vspace{-4mm}
\end{figure}

RRAM is another resistance change memory that relies on microstructural change in the material that causes the cell to have two resistance states (LRS and HRS). As shown in Fig.~\ref{fig:RRAM}, the RRAM cell consists of a metal-oxide-metal (MOM) stack in which the sub-oxide is typically $\text{TaO}_\text{x}$, $\text{HfO}_\text{x}$ or $\text{TiO}_\text{x}$. The devices do not start off as being resistive switching memories; they have to go through a one-time programming process known as \emph{forming}. The forming process involves the application of a high voltage pulse that causes the oxide to breakdown and form a conductive filament that shunts the two metal electrodes, causing the resistance to decrease~\cite{Sharma2014RRAM}. The LRS corresponds to the shunted conductive filament. This filament can be disconnected by applying a voltage of the opposite polarity. Once the conductive filament is disconnected, the device resistance increases, and the device is said to be in the HRS. The device can now be cycled between LRS and HRS by applying voltages of opposite polarity as shown in Fig.~\ref{fig:RRAM}.

As the RRAM switching mechanism is filamentary in nature, the RRAM devices are highly scalable, operate at ultra-low powers, have good retention characteristics, and can be integrated into a compact crossbar array~\cite{Wong2012RRAM}. 

However, similar to PCM, RRAM also suffers from limited endurance, especially when operated at low power~\cite{Chen2013endurance}. In RRAM, the stuck-at defects may be additionally introduced during the forming process due to poor power-limiting during the breakdown~\cite{Sharma2014RRAM}.

The stuck-at LRS defects in RRAM have been attributed to the widening of the conductive filament~\cite{Yalon2015}. Once the filament widens, the device resistance drops and the RESET power is insufficient to disconnect the filament. This causes the cell to be permanently set to LRS. The widening of the filament is thought of as a stochastic increase in the number of oxygen vacancies in the filament during the SET and forming operation. It can be explained by an incomplete retraction of oxygen vacancies during the previous RESET~\cite{Kwon2015}. Similarly, the devices can also suffer from a stuck-at-HRS defect if the devices undergo over-RESET~\cite{Lee2010RRAM}. In this process, the oxygen vacancies are retracted irreversibly, making the device stuck-at HRS defect. 

Similar to PCM, once the device starts experiencing the over-SET or over-RESET which precedes endurance failure, the devices would undergo a positive feedback that would make the stuck-at defects imminent. Moreover, as the endurance failure is mediated by stochastic motion of oxygen vacancies during the SET or RESET processes~\cite{Chen2015endurance}, it is very difficult to prevent these stuck-at defects.

\section{Channel Model}\label{sec:channelmodel}

In Section~\ref{sec:background}, we explained that both PCM and RRAM suffer from stuck-at HRS or LRS defects. The resistance state can be sensed as either 0 or 1, depending on the sensing scheme of read operation (e.g., HRS $\rightarrow$ 0, LRS $\rightarrow$ 1 or vice versa). Thus, we can claim that the defect channel model by Kuznetsov and Tsybakov~\cite{Kuznetsov1974} is a proper mathematical model for resistive memories. After providing notation, we will explain the defect channel model. 

\subsection{Notation}

We use parentheses to construct column vectors from comma separated lists. For a $n$-tuple column vector $\mathbf{a} \in \mathbb{F}_q^n$ (where $\mathbb{F}_q$ denotes the finite field with $q$ elements and $\mathbb{F}_q^n$ denotes the set of all $n$-tuple vectors over $\mathbb{F}_q$), we have
\begin{equation} \label{eq:vector}
	(a_1, \ldots, a_n) = \begin{bmatrix}
	a_1 \\ \vdots \\ a_n
	\end{bmatrix} = \left[a_1 \: \ldots \: a_n \right]^T
\end{equation}
where superscript $T$ denotes transpose. Note that $a_i$ represents the $i$-th element of $\mathbf{a}$. For a binary vector $\mathbf{a} \in \mathbb{F}_2^n$, $\overline{\mathbf{a}}$ denotes the bit-wise complement of $\mathbf{a}$. For example, the $n$-tuple all-ones vector $\mathbf{1}_n$ is equal to $\overline{\mathbf{0}}_n$ where $\mathbf{0}_n$ is the $n$-tuple all-zero vector. Also, $\mathbf{0}_{m, n}$ denotes the $m \times n$ all-zero matrix. 

In addition, $\| \mathbf{a} \|$ denotes the Hamming weight of $\mathbf{a}$ and $\text{supp}(\mathbf{a})$ denotes the support of $\mathbf{a}$. We use the notation of $[i:j] = \left\{i, i+1, \ldots, j\right\}$ for $i < j$ and $[n] = [1:n]= \left\{1, \ldots, n \right\}$. Note that $\mathbf{a}_{[i:j]} = \left(a_i, \ldots, a_j\right)$ and $\mathbf{a}_{\setminus i} = \left(a_1, \ldots, a_{i-1}, a_{i+1}, \ldots, a_n\right)$.

\subsection{Channel Model: Defect Channel}

\begin{figure}[!t]
	\centering
	\includegraphics[width=0.33\textwidth]{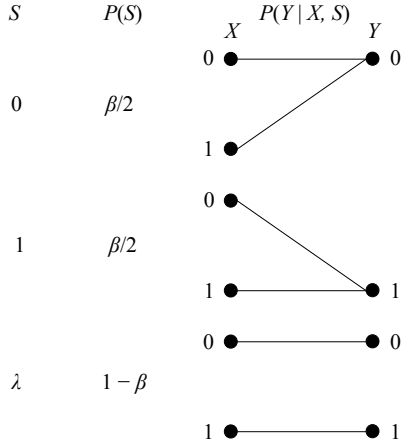}
	\caption{Binary defect channel.}
	\label{fig:BDC}
	\vspace{-3mm}
\end{figure}

We summarize the defect channel model in~\cite{Kuznetsov1974}. Define a variable $\lambda$ that indicates whether the memory cell is defective or not and $\widetilde{\mathbb{F}}_q = \mathbb{F}_q \cup \{\lambda \}$. Let ``$\circ$'' denote the operator  $\circ:\mathbb{F}_q \times \widetilde{\mathbb{F}}_q \rightarrow \mathbb{F}_q$ as in~\cite{Heegard1983plbc}
\begin{equation}\label{eq:circ_operator}
x \circ s =
\begin{cases}
x, & \text{if } s = \lambda ; \\
s, & \text{if } s \ne \lambda.
\end{cases}
\end{equation}
By using the operator $\circ$, an $n$-cell memory with defects is modeled by
\begin{equation}
\mathbf{y} = \mathbf{x} \circ \mathbf{s} \label{eq:BDC_vector}
\end{equation}
where $\mathbf{x}, \mathbf{y}\in \mathbb{F}_q^n$ are the channel input and output vectors. Also, the channel state vector $\mathbf{s} \in \widetilde{\mathbb{F}}_q^n$ represents the defect information in the $n$-cell memory. Note that $\circ$ is the vector component-wise operator. 

If $s_i = \lambda$, this $i$-th cell is called \emph{normal}. If the $i$-th cell is \emph{defective} (i.e., $s_i \ne \lambda$), its output $y_i$ is stuck-at $s_i$ independent of the input $x_i$. So, the $i$-th cell is called stuck-at defect whose stuck-at value is $s_i$. The probabilities of stuck-at defects and normal cells are given by
\begin{equation}\label{eq:defect_probability}
	P(S = s) =
	\begin{cases}
		1 - \beta, & \text{if } s = \lambda ; \\
		\frac{\beta}{q}, & \text{if } s \ne \lambda
	\end{cases}
\end{equation}
where the probability of stuck-at defects is $\beta$. Fig.~\ref{fig:BDC} shows the binary defect channel for $q=2$. 

In the defect channel model, it is assumed that the encoder knows the side information of defects before writing data to memories~\cite{Kuznetsov1974}. Hence, it can be explained by Gelfand-Pinsker problem~\cite{Gelfand1980}.  

\section{Locally Rewritable Codes (LWC)}\label{sec:lwc}

\subsection{Motivation and Toy Example}

As a toy example, suppose that $n$-cell \emph{binary} memory has a single stuck-at defect. It is easy to see that this stuck-at defect can be handled by the following simple technique~\cite{Kuznetsov1974}. 
\begin{equation} \label{eq:single_defect}
	\mathbf{c} = ( \mathbf{m}, 0) + \mathbf{1}_n \cdot p
\end{equation}
where $\mathbf{c}\in\mathbb{F}_2^n$ is a codeword and $\mathbf{m}\in\mathbb{F}_2^k$ is an information (message) where $k = n-1$ and $p$ is a parity (redundant) bit. 

Suppose that $i$-th cell is a defect whose stuck-at value is $s_i \in \mathbb{F}_2$. If $i \in [n-1]$ and $s_i = m_i$, or if $i=n$ and $s_n = 0$, then $p$ should be 0. Otherwise, $p=1$. Thus, $p$ decides whether to flip $\mathbf{m}$ or not. It is worth mentioning that this simple coding is optimal since it achieves the following upper bound in \cite{Kuznetsov1974} with equality. 
\begin{equation} \label{eq:bound_defect}
n - t - \left\lceil \log_2 \ln 2^t \binom{n}{t} \right\rceil \le \log_2 \mathcal{M} \le n - t
\end{equation}
where $\mathcal{M}$ is the number of codewords and $t$ is the number of stuck-at defects among $n$ cells. For linear codes, $k = \log_2 \mathcal{M}$, i.e., $k \le n-1$. 

If there is no stuck-at defect among $n$ cells, then we can store $\mathbf{m}$ by writing $\mathbf{c} = (\mathbf{m}, 0)$ (i.e., $p=0$). Now, consider the case when stored information needs to be updated causing $\mathbf{m}$ to become $\mathbf{m}'$. Usually, $\| \mathbf{m}- \mathbf{m}' \| \ll n$, which happens often due to the updates of files. Instead of storing $\mathbf{m}'$ into another group of $n$ cells, it is more efficient to store $\mathbf{m}'$ by rewriting only  $\| \mathbf{m}- \mathbf{m}' \|$ cells. For example, suppose that $m_i' \ne m_i$ for an $i \in [k]$ and $m_j' = m_j$ for all other $j \in [k] \setminus i$. Then, we can store $k$-bit $\mathbf{m}'$ by rewriting only $i$-th cell. 

An interesting problem arises when a cell to be rewritten is defective. Suppose that $i$-th cell is a stuck-at defect whose stuck-at value is $s_i$. If $s_i = m_i \ne m_i'$, then we should write $\mathbf{c} = (\mathbf{m}, 0)$ for storing $\mathbf{m}$. However, in order to store the updated information $\mathbf{m}'$, we should write $\mathbf{c}' = \overline{\mathbf{c}} = (\overline{\mathbf{m}}, 1)$ where $p=1$. Thus, $n-1$ cells should be rewritten to update one bit data $m_i'$ without stuck-at error. The same thing happens when $s_i = m_i' \ne m_i $. When considering endurance limit and power consumption, rewriting $n-1$ cells is a high price to pay for preventing one bit stuck-at error.  

In order to relieve this burden, we change \eqref{eq:single_defect} by introducing an additional parity bit as follows. 
\begin{align} \label{eq:single_defect_locality}
	\mathbf{c} & = \left( \mathbf{m}_{[1:\frac{n}{2}]}, 0,\mathbf{m}_{[\frac{n}{2}+1:n]} , 0 \right) + G_0 \mathbf{p} \\
	&= \left( \mathbf{m}_{[1:\frac{n}{2}]}, 0,\mathbf{m}_{[\frac{n}{2}+1:n]} , 0 \right) + \begin{bmatrix} \mathbf{1}_{\frac{n}{2}} & \mathbf{0}_{\frac{n}{2}} \\ \mathbf{0}_{\frac{n}{2}} & \mathbf{1}_{\frac{n}{2}} \end{bmatrix} (p_1, p_2) 
\end{align}
where $k = n-2$. For simplicity's sake, we assume that $n$ is even. Then, $\mathbf{1}_{\frac{n}{2}}$ and $\mathbf{0}_{\frac{n}{2}}$ are all-ones and all-zeros column vectors with ${n}/{2}$ elements. By introducing an additional parity bit, we can reduce the number of rewriting cells from $n-1$ to $\frac{n}{2}-1$. 

This idea is similar to the concept of Pyramid codes which are the early LRC~\cite{Huang2007pyramid}. For $n$ disk nodes, single parity check codes can repair one node failure (i.e., single erasure) by 
\begin{equation} \label{eq:single_erasure}
\mathbf{1}_n^T \widehat{\mathbf{c}} = 0
\end{equation} 
where $\widehat{\mathbf{c}}$ represents the recovered codeword from disk node failures. Assuming that $c_i$ is erased due to a node failure, $c_i$ can be recovered by 
\begin{equation} \label{eq:single_erasure_repair}
\widehat{c}_i = c_i = \sum_{j \in [n]\setminus i}{c_j}.
\end{equation}
For this recovery, we should access $k = n-1$ nodes which degrades the repair speed. For more efficient repair process, we can add a new parity as follows. 
\begin{equation} \label{eq:single_erasure_locality}
H^T \widehat{\mathbf{c}} = \begin{bmatrix} \mathbf{1}_{\frac{n}{2}} & \mathbf{0}_{\frac{n}{2}} \\ \mathbf{0}_{\frac{n}{2}} & \mathbf{1}_{\frac{n}{2}} \end{bmatrix}^T \widehat{\mathbf{c}} = \mathbf{0}
\end{equation} 
Then, a failed node $c_i$ can be repaired by accessing only $\frac{n}{2} - 1$ nodes. Note that the repair locality of \eqref{eq:single_erasure_locality} is $\frac{n}{2}-1$ whereas the repair locality of \eqref{eq:single_erasure} is $n - 1$ which is a simple but effecitve idea of Pyramid codes.   

An interesting observation is that $G_0$ of \eqref{eq:single_defect_locality} is the same as $H$ of \eqref{eq:single_erasure_locality}. In addition, note that the number of resistive memory cells to be rewritten is the same as the number of nodes to be accessed in distributed storage systems. This observation will be further discussed in Subsection~\ref{subsec:duality}. 

\subsection{Locally Rewritable Codes}

In this subsection, we propose LWC by generalizing the idea of the toy example in the previous subsection. A traditional coding scheme for  defect channel is \emph{additive encoding} which masks defects by adding a carefully selected
vector. The goal of masking stuck-at defects is
to make a codeword whose values at the locations of defects
match the stuck-at values of corresponding  defects \cite{Kuznetsov1974, Tsybakov1975additive}. The additive encoding can be formulated as
\begin{equation}\label{eq:additive_encoding}
\mathbf{c} = (\mathbf{m}, \mathbf{0}_{n-k}) + \mathbf{c}_0 = (\mathbf{m}, \mathbf{0}_{n-k}) + G_0 \mathbf{p}
\end{equation}
where $G_0 \in \mathbb{F}_q^{n \times (n-k)}$. By adding a vector $\mathbf{c}_0 = G_0 \mathbf{p} \in \mathcal{C}_0$, we can mask stuck-at defects among $n$ cells. For the systematic codes, $G_0$ is given by~\cite{Heegard1983plbc}
\begin{equation}
G_0 = \begin{bmatrix} R \\ I_{n-k} \end{bmatrix}
\end{equation}
where $R \in \mathbb{F}_2^{k \times (n-k)}$ and $I_{n-k}$ is the $(n-k)$-dimensional identity matrix. Note that the identity matrix is located in the parity part unlike the conventional error-control codes. 

The decoding can be given by
\begin{equation} \label{eq:decoding}
\widehat{\mathbf{m}} = H_0^T \mathbf{y}
\end{equation}
where $\widehat{\mathbf{m}}$ represents the recovered message of $\mathbf{m}$. Note that the parity check matrix $H_0$ of $\mathcal{C}_0$ is given by $H_0 = [I_k \quad R]^T$ and $H_0^T G_0 = \mathbf{0}_{k, n-k}$. Note that \eqref{eq:decoding} is equivalent to the equation of coset codes. 

The \emph{minimum distance} of additive encoding is given by 
\begin{align} \label{eq:dmin}
d^{\star} &= \underset{
\substack{
	\mathbf{x} \ne \mathbf{0} \\
	G_0^T \mathbf{x}= \mathbf{0}
}}
{\text{min }} \|\mathbf{x}\|
\end{align}
which means that any $d^{\star} - 1$ rows of $G_0$ are linearly independent. Thus, additive encoding guarantees masking up to $d^{\star}-1$ stuck-at defects~\cite{Tsybakov1975additive, Heegard1983plbc}. 

Now we investigate \emph{rewriting locality} of additive encoding. As repair locality of LRC is meaningful only for single disk failure, rewriting locality is valid when there is one stuck-at defect among $n$ cells. In distributed storage systems, the most common case is a single node failure among $n$ nodes~\cite{Huang2007pyramid}. Similarly, for a proper defect probability $\beta$, we can claim that the most common scenario of resistive memories is that there is a single stuck-at defect among $n$ cells. 

We define \emph{initial writing cost} and \emph{rewriting cost} which are related to write endurance and power consumption. 

\begin{definition} [Initial Writing Cost] Suppose that $\mathbf{m}$ was stored by its codeword $\mathbf{c}$ in the initial stage of $n$ cells where all the normal cells are set to zeros. The writing cost is given by
	\begin{equation} \label{eq:writing_cost_def}
	\Delta (\mathbf{m}) = \| \mathbf{c}  \| - t_{\setminus 0}
	\end{equation}
	where $t_{\setminus 0}$ denotes the number of stuck-at defects whose stuck-at values are nonzero. 
\end{definition}

In \eqref{eq:writing_cost_def}, we assume that there are $t$ stuck-at defects among $n$ cells and $\mathbf{c}$ masks these $t$ stuck-at defects successfully. So, we do not need to write stuck-at defects since their stuck-at values are the same as corresponding elements of $\mathbf{c}$. 

\begin{definition} [Rewriting Cost] Suppose that $\mathbf{m}$ was stored by its codeword $\mathbf{c}$ in $n$ cells. If $\mathbf{c}'$ is rewritten to these $n$ cells to store the updated $\mathbf{m}'$, the rewriting cost is given by
\begin{equation} \label{eq:rewriting_cost_def}
\Delta (\mathbf{m}, \mathbf{m}') = \| \mathbf{c} - \mathbf{c}' \| 
\end{equation}
where we assume that both $\mathbf{c}$ and $\mathbf{c}'$ mask stuck-at defects. 
\end{definition}

High rewriting cost implies that the states of lots of cells should be changed, which is harmful to write endurance and increases power consumption. 

It is worth mentioning that, in general, the rewriting cost is more important than the initial writing cost since most of write operations will be rewriting. If a device offers write endurance of 10000 cycles, the write operations of 9999 will be rewriting whereas only one among 10000 writing is the initial write operation (i.e., 0.01\%). However, there may be some storage applications (such as for archival storage), where the number of initial writings and rewritings may be similar.

Now, we introduce the \emph{rewriting locality} which affects initial writing cost and rewriting cost.
\begin{definition} [Information Rewriting Locality] \label{def:inf_r} Suppose that $m_i$ for $i \in [k]$, i.e., information (message) part, should be updated to $m_i' \ne m_i$ and the corresponding $i$-th cell is a stuck-at defect. If $m_i$ can be updated to $m_i'$ by rewriting $r^{\star}$ other cells, then the $i$-th coordinate has \emph{information rewriting locality} $r^{\star}$. 
\end{definition}

\begin{lemma} \label{thm:inf_r} If the $i$-th coordinate for $i \in [k]$ has information rewriting locality $r^{\star}$, then there exists $\mathbf{c}_0 \in \mathcal{C}_0$ such that $i \in \text{supp}(\mathbf{c}_0)$ and $\left\| \mathbf{c}_0 \right\| = r^{\star}+1$.
\end{lemma}
\begin{IEEEproof}
	For $\mathbf{m}$ and $\mathbf{m}'$, suppose that $m_i \ne m_i'$ for an $i \in [k]$ and $m_j = m_j'$ for all other $j \in [k] \setminus i$. Note that $i$-th cell is a stuck-at defect whose stuck-at value is $s_i$. We should consider the following cases:
	\begin{enumerate}
		\item 	$m_i' \ne m_i = s_i$. 
		\item 	$m_i \ne m_i' = s_i$. 
		\item 	$m_i \ne m_i'$, $m_i \ne s_i$ and $m_i' \ne s_i$. 
	\end{enumerate}	
	
	For	$m_i' \ne m_i = s_i$, it is obvious that $\mathbf{c} = (\mathbf{m}, \mathbf{0}_{n-k})$ and $\mathbf{c}' = (\mathbf{m}', \mathbf{0}_{n-k})  + \mathbf{c}_0'$ where $c_i' = m_i' + c_{0, i}' = s_i$ by~\eqref{eq:additive_encoding}. For the information rewriting locality $r^{\star}$, $\mathbf{c}_0' \in \mathcal{C}_0$ should satisfy $\left\| \mathbf{c}_0' \right\| = r^{\star}+1$ to mask the stuck-at defect by writing $r^{\star}$ cells. Note that we do not need to write the stuck-at defect since its stuck-at value is $s_i = c_i'$. For $m_i \ne m_i' = s_i$, the proof is similar. 
	
	For $m_i \ne m_i'$, $m_i \ne s_i$ and $m_i' \ne s_i$, $\mathbf{c} = (\mathbf{m}, \mathbf{0}_{n-k}) + \mathbf{c}_0$ and $\mathbf{c}' = (\mathbf{m}', \mathbf{0}_{n-k})  + \mathbf{c}_0'$. We can pick $\mathbf{c}_0$ and $\mathbf{c}_0'$ such that $\mathbf{c}_0' = \alpha \mathbf{c}_0$ (where $\alpha \in \mathbb{F}_q $) and $m_i + c_{0, i} = m_i' + c_{0, i}' = s_i$. For the information rewriting locality $r^{\star}$, $\mathbf{c}_0$ and $\mathbf{c}_0'$ should satisfy $\left\| \mathbf{c}_0 \right\| = \left\| \mathbf{c}_0' \right\| = r^{\star} + 1$. 	
\end{IEEEproof}

%If a stuck-at defect's coordinate is $i \in [k+1:n]$, i.e. parity location, then $\mathbf{m}$ can be updated to $\mathbf{m}'$ by rewriting $\| \mathbf{m} - \mathbf{m}' \|$ cells because of $\mathbf{c}_0 = \mathbf{c}_0'$. Thus, a stuck-at defect in the parity location is not related to rewriting. However, a stuck-at defect in the parity location affects initial writing. We will define parity rewriting locality as follows.   

\begin{definition} [Parity Rewriting Locality] \label{def:par_r} Suppose that only one nonzero symbol $m_i$ should be stored to the initial stage of $n$ cells. Note that there is a stuck-at defect in the parity location $j$ for $j \in [k+1:n]$ (i.e., parity part) and $s_j \ne 0$. If $m_i$ can be stored by writing at most $r^{\star} + 1$ cells, then the $j$-th coordinate has \emph{parity rewriting locality} $r^{\star}$. 
\end{definition}

\begin{lemma} \label{thm:par_r} If the $j$-th coordinate for $j \in [k+1:n]$ has parity rewriting locality $r^{\star}$, then there exists $\mathbf{c}_0 \in \mathcal{C}_0$ such that $j \in \text{supp} (\mathbf{c}_0)$ and $\left\| \mathbf{c}_0 \right\| = r^{\star}+1$.
\end{lemma}
\begin{IEEEproof}
Suppose that $\mathbf{m}$ should be stored to the initial stage of $n$ cells where $m_i \ne 0$ for $i \in [k]$ and $m_i' = 0$ for $i' \in [k] \setminus i$. By~\eqref{eq:additive_encoding}, $\mathbf{c} = (\mathbf{m}, \mathbf{0}_{n-k}) + \mathbf{c}_0 $ such that $c_{0, j} = s_j$ where the $j$-th cell is a stuck-at defect for $j \in [k+1:n]$. For the parity rewriting locality $r^{\star}$, $\mathbf{c}_0$ should satisfy $\left\| \mathbf{c}_0 \right\| = r^{\star}+1$. If $i \in \text{supp}(\mathbf{c}_0)$, then it is possible to store $m_i$ without stuck-at error by writing $\|\mathbf{c}_{0 \setminus j} \| = r^\star$ cells since we do not need to write $c_{0, j}$. Otherwise, we should write both $m_i$ and $\|\mathbf{c}_{0 \setminus j} \|$, i.e., $r^\star + 1$ cells. 	
\end{IEEEproof}

\begin{definition} [Locally Rewritable Codes] If any $i$-th coordinate for $i \in [n]$ has (information or parity) rewriting locality at most $r^{\star}$, then this code is called locally rewritable code (LWC) with rewriting locality $r^{\star}$. $(n, k, d^{\star}, r^{\star})$ LWC code is a code of length $n$ with information length $k$, minimum distance $d^{\star}$, and rewriting locality $r^{\star}$.
\end{definition}

Now, we show in the following theorem that rewriting locality $r^{\star}$ is an important parameter for rewriting cost. 

\begin{theorem} \label{thm:rewriting_cost} Suppose that $\mathbf{m}$ is updated to $\mathbf{m}'$ by LWC with rewriting locality $r^{\star}$. If there is a single stuck-at defect in $n$ cells, then the rewriting cost $\Delta (\mathbf{m}, \mathbf{m}')$ is given by
\begin{equation} \label{eq:rewriting_cost_bound}
\Delta (\mathbf{m}, \mathbf{m}') \le \|\mathbf{m} - \mathbf{m}' \| + r^{\star} - 1.
\end{equation}
\end{theorem}
\begin{IEEEproof} First, suppose that the single defect's coordinate is $i \in [k]$ and its stuck-at value is $s_i$. If $m_i = m_i'$, then $\Delta (\mathbf{m}, \mathbf{m}') = \| \mathbf{m} - \mathbf{m}' \|$ since $\mathbf{c} = (\mathbf{m}, \mathbf{0}) + \mathbf{c}_0$ and $\mathbf{c}' = (\mathbf{m}', \mathbf{0}) + \mathbf{c}_0' = (\mathbf{m}', \mathbf{0}) + \mathbf{c}_0$ where $\mathbf{c}_0' =\mathbf{c}_0$. 
	
If $m_i \ne m_i' = s_i$, then $\mathbf{c} = (\mathbf{m}, \mathbf{0}) + \mathbf{c}_0 $ and $\mathbf{c}' = (\mathbf{m}', \mathbf{0}) + \mathbf{c}_0' = (\mathbf{m}', \mathbf{0})$ where $\mathbf{c}_0' = \mathbf{0}$. In order to mask the stuck-at defect, $i \in \text{supp}(\mathbf{c}_0)$.  Then,
\begin{align}
\Delta (\mathbf{m}, \mathbf{m}') &= \| \mathbf{c} - \mathbf{c}' \| = \| (\mathbf{m}, \mathbf{0}) + \mathbf{c}_0 - (\mathbf{m}', \mathbf{0}) \| \\
& = \| (\mathbf{m}-\mathbf{m}', \mathbf{0}) + \mathbf{c}_0 \| \\
& \le \| \mathbf{m}-\mathbf{m}' \| +  \|\mathbf{c}_0 \| - 2 \label{eq:rewriting_cost_pf1} \\
& = \| \mathbf{m}-\mathbf{m}' \| +  r^{\star} - 1 \label{eq:rewriting_cost_pf2}
\end{align}
where \eqref{eq:rewriting_cost_pf1} follows from the  $m_i - m_i' = c_{0,i}$ where $c_{0,i}$ is the $i$-th element of $\mathbf{c}_0$. Also, \eqref{eq:rewriting_cost_pf2} follows from Lemma~\ref{thm:inf_r}. By supposing $m_i \ne m_i'$ for $i \in [k]$ and $m_j = m_j'$ for $j \in [k]\setminus i$, i.e., $\| \mathbf{m}-\mathbf{m}' \| = 1$, we confirm that the RHS of \eqref{eq:rewriting_cost_pf2} is $r^{\star}$, which coincides with Definition~\ref{def:inf_r}. For $m_i' \ne m_i = s_i$, we can show \eqref{eq:rewriting_cost_bound} by a similar method. 

Next, suppse that the single defect's coordinate is $i \in [k+1:n]$. Since  $\mathbf{c} = (\mathbf{m}, \mathbf{0}) + \mathbf{c}_0$ and $\mathbf{c}' = (\mathbf{m}', \mathbf{0}) + \mathbf{c}_0$, the rewriting cost is $\Delta (\mathbf{m}, \mathbf{m}') = \| \mathbf{m}-\mathbf{m}' \|$. 
\end{IEEEproof}

\begin{corollary} \label{thm:writing_cost}If $\mathbf{m}$ is stored in the initial stage of $n$ cells with a single stuck-at defect, then the writing cost $\Delta (\mathbf{m})$ is given by
\begin{equation} \label{eq:writing_cost}
\Delta (\mathbf{m}) \le \|\mathbf{m} \| + r^{\star}.
\end{equation}
\end{corollary}
\begin{IEEEproof}
First suppose that the single defect's coordinate is $i \in [k]$ and its stuck-at value is $s_i$. If $s_i = m_i$, then $\mathbf{c} = (\mathbf{m}, \mathbf{0})$. The writing cost $\Delta (\mathbf{m}) = \| \mathbf{c} \| - 1 = \| \mathbf{m} \| - 1$. 

If $s_i \ne m_i$, then $\mathbf{c} = (\mathbf{m}, \mathbf{0}) + \mathbf{c}_0$. Then, 
\begin{align} 
\Delta (\mathbf{m}) &= \|\mathbf{c} \| - 1 \\
& \le  \|\mathbf{m} \| + \| \mathbf{c}_0 \| - 1 \label{eq:writing_cost_pf0}\\
& =  \|\mathbf{m} \| + r^{\star} \label{eq:writing_cost_pf1}
\end{align}
where \eqref{eq:writing_cost_pf1} follows from Lemma~\ref{thm:inf_r}. 

Next suppose that the single defect's coordinate is $j \in [k+1: n]$. If $s_j = 0$, then $\Delta (\mathbf{m}) = \|\mathbf{c}\| =\|(\mathbf{m}, \mathbf{0}) \| =  \|\mathbf{m}\|$. If $s_j \ne 0$, then $\mathbf{c} = (\mathbf{m}, \mathbf{0}) + \mathbf{c}_0$ where $j \in \text{supp}(\mathbf{c}_0)$. By Lemma~\ref{thm:par_r}, we can claim that $\Delta (\mathbf{m}) \le \|\mathbf{m} \| + \| \mathbf{c}_0 \| - 1 = \|\mathbf{m} \| + r^{\star} $. 
\end{IEEEproof}

%In resistive memories, we can claim that rewriting cost is more important than initial writing cost because there is no erase operation as in flash memories. All the initial stage of cells would be used up after a while, hence most of data will be stored by rewriting resistive memory cells. 

Theorem~\ref{thm:rewriting_cost} and Corollary~\ref{thm:writing_cost} show that a small rewriting locality $r^{*}$ can reduce the writing cost and rewriting cost, which is helpful for improving endurance and power consumption. 

\subsection{Duality of LRC and LWC} \label{subsec:duality}

In this subsection, we investigate the duality of LRC and LWC. We show that existing construction methods of LRC can be used to construct LWC based on this duality. First, the relation between minimum distance $d^{\star}$ and rewriting locality $r^{\star}$ is observed. 

%\begin{claim} \label{thm:dist_r}Let $\mathcal{C}_0$ denote a linear code whose generator matrix is $G_0$ of \eqref{eq:additive_encoding}. If there exists $\mathbf{c}_0 \in \mathcal{C}_0$ such that $i \in \text{supp}\left(\mathbf{c}_0 \right)$ and $\| \mathbf{c}_0 \| = r^{\star} + 1$, then the $i$-th element has rewriting locality $r^{\star}$.   
%\end{claim}
%\begin{IEEEproof}
%While proving Theorem~\ref{thm:rewriting_cost} and Corollary~\ref{thm:writing_cost}, we use the condition of $\| \mathbf{c}_0 \| - 1 = r^{\star}$ where the $i$-th coordinate belongs to the support of $\mathbf{c}_0$. 
%\end{IEEEproof}	

\begin{definition} \label{def:cyc_LWC}If $\mathcal{C}_0$ is cyclic, then the LWC is called \emph{cyclic}. 
\end{definition}

\begin{lemma}\label{thm:cyc}Let $\mathcal{C}_0$ denote a cyclic code whose minimum distance is $d_0$. Then, corresponding cyclic LWC's rewriting locality is $r^{\star} = d_0 - 1$.	
\end{lemma}
\begin{IEEEproof}
Due to the property of cyclic codes, we can claim that there exists $\mathbf{c}_0 \in \mathcal{C}_0$ such that $i \in \text{supp}(\mathbf{c}_0)$ and $\| \mathbf{c}_0 \| = d_0$ for any $i \in [n]$. Since $d_0$ is the minimum distance of $\mathcal{C}_0$, we can claim that the rewriting locality is $r^{\star} = d_0 - 1$. 
\end{IEEEproof}

From the definition of $d^{\star}$ in \eqref{eq:dmin}, $d^{\star} = d_0^{\perp}$ which is the minimum distance of $\mathcal{C}_0^{\perp}$, namely, dual code of $\mathcal{C}_0$. Thus, the parameters of cyclic LWC is given by
\begin{equation} \label{eq:LWC_parameter}
(d^{\star}, r^{\star}) = (d_0^{\perp}, d_0 - 1). 
\end{equation}
In \cite{Huang2015, Tamo2015subcodes}, an equivalent relation for cyclic LRC was given by
\begin{equation} \label{eq:LRC_parameter}
(d, r) = (d, d^{\perp}-1).
\end{equation}

By comparing \eqref{eq:LWC_parameter} and \eqref{eq:LRC_parameter}, we observed the \emph{duality} between LRC and LWC. This duality is important since it indicates that we can construct LWC using existing construction methods of LRC as shown in the following theorem. 

\begin{theorem} \label{thm:LWC_construction}
Suppose that $H_{\text{LRC}} \in \mathbb{F}_q^{n \times (n-k)}$ is the parity check matrix of cyclic LRC $\mathcal{C}_{\text{LRC}}$ with $(d, r) = (d, d^{\perp} - 1)$. By setting $G_0 = H_{\text{LRC}}$, we can construct cyclic LWC $\mathcal{C}_{\text{LWC}}$ with 
\begin{equation}
(d^{\star}, r^{\star}) = (d, d^{\perp} - 1).
\end{equation}
\end{theorem}
\begin{IEEEproof}
By setting $G_0 = H_{\text{LRC}}$, the LWC's codeword $\mathbf{c} \in \mathcal{C}_{\text{LWC}}$ is given by
\begin{equation*}\label{eq:additive_encoding_dual}
\mathbf{c} = (\mathbf{m}, \mathbf{0}) + H_{\text{LRC}} \cdot \mathbf{p}. 
\end{equation*}
The minimum distance $d^{\star}$ of $\mathcal{C}_{\text{LWC}}$ is given by
\begin{align*} \label{eq:dmin_dual}
d^{\star} &= \underset{
	\substack{
		\mathbf{x} \ne \mathbf{0} \\
		H_{\text{LRC}}^T \mathbf{x}= \mathbf{0}
	}}
	{\text{min }} \|\mathbf{x}\|
\end{align*}
which is equivalent to the minimum distance $d$ of $\mathcal{C}_{\text{LRC}}$. Hence, we can claim that
\begin{equation} \label{eq:dmin_relation}
d^{\star}=d_0^{\perp} = d.
\end{equation} 
From \eqref{eq:LWC_parameter} and \eqref{eq:dmin_relation}, $r^{\star}=d_0 - 1 = d^{\perp} - 1$.
\end{IEEEproof}

\begin{remark}[Optimal Cyclic LWC] \label{thm:LWC_optimal}
	Theorem~\ref{thm:LWC_construction} shows that the optimal cyclic $(n, k, r, d)$ LRC can be used to construct the optimal cylic $(n, k, r^\star, d^\star)$ LWC such that
	\begin{equation}
	d^{\star} = n - k - \left \lceil \frac{k}{r^{\star}} \right \rceil + 2.
	\end{equation} 
\end{remark}
Hence, the optimal LWC can be constructed from the optimal LRC. 

\begin{remark}[Bound of LWC] 
	From Theorem~\ref{thm:LWC_construction} and Remark~\ref{thm:LWC_optimal}, we can claim the following bound for LWC. 
	\begin{equation}
		d^{\star} \le n - k - \left \lceil \frac{k}{r^{\star}} \right \rceil + 2
	\end{equation}
	which is equivalent to the bound for LRC given by \eqref{eq:general_singleton}. 		
\end{remark}

\begin{table}[t]
	\renewcommand{\arraystretch}{1.4}
	\caption{Duality of LRC and LWC}
	\label{tab:duality}
	\centering
	{\hfill{}
	\begin{tabular}{|c|c|c|}
		\hline
		 		    	& $(n, k, d, r)$ LRC    & $(n, k, d^{\star}, r^{\star})$ LWC \\ \hline \hline
		\multirow{2}{*}{Application}     & Distributed storage systems & Resistive memories \\ 
										 & (system level)              & (physical level) \\ \hline
		Channel   & Erasure channel             & Defect channel   \\ \hline
		Encoding        & $\mathbf{c} = G_{\text{LRC}}\mathbf{m}$ & $\mathbf{c} = (\mathbf{m}, \mathbf{0})+ H_{\text{LRC}}\mathbf{p}$   \\ \hline
		Decoding        & $H_{\text{LRC}}^T \widehat{\mathbf{c}} = \mathbf{0}$ & $G_{\text{LRC}}^T \mathbf{c} = \widehat{\mathbf{m}}$  \\ \hline
		Bound           & $d \le n - k - \left \lceil \frac{k}{r} \right \rceil + 2$ & $d^{\star} \le n - k - \left \lceil \frac{k}{r^{\star}} \right \rceil + 2$ \\ \hline
		\multirow{2}{*}{Trade-off} & $d$ (reliability) vs. & $d^{\star}$ (reliability) vs. \\
		& $r$ (repair efficiency) & $r^*$ (rewriting cost)  \\ \hline
	\end{tabular}}
	\hfill{}
	\vspace{-3mm}
\end{table}

In Table~\ref{tab:duality}, the duality properties of LRC and LWC are summarized. In the decoding of LRC, $\widehat{\mathbf{c}}$ denotes a recovered codeword from node failures. In addition, $\widehat{\mathbf{m}}$ in the decoding of LWC represents the recovered message. It is worth mentioning that this duality can be connected to the duality between erasures and defects~\cite{Kim2014duality}.

%\begin{figure}[!t]
%\centering
%\subfloat[Single-level cell (SLC) for $B=1$]{\includegraphics[width=3in]{fig_slc.eps}
%\label{fig:slc}}
%\hfil
%\vspace{-1mm}
%\subfloat[Multi-level cell (MLC) for $B=2$]{\includegraphics[width=3in]{fig_mlc.eps}
%\label{fig:mlc}}
%\caption{Threshold voltage distribution of flash memory cells.}
%\label{fig:slc_mlc}
%\vspace{-5mm}
%\end{figure}

%\begin{figure}[t]
%   \centering
%   \includegraphics[width=0.29\textwidth]{fig_defect_channel.eps}
%   \caption{Channel model of a memory with defective cells.}
%   \label{fig:defect_channel}
%   \vspace{-5mm}
%\end{figure}

\section{Conclusion and Future Work}\label{sec:conclusion}

Inspired by LRC for distributed storage systems, we proposed LWC to improve endurance limit and power consumption of resistive memories. We showed the relation between rewriting cost and rewriting locality of LWC. Also, we investigated the duality between LRC and LWC, which makes it possible to construct LWC by using existing construction methods of LRC. 

As part of our future work, we plan to evaluate the performance of LWC. Although some recent works have investigated the characteristics of endurance limit~\cite{Chen2013endurance, Chen2015endurance}, a proper channel model is not available.\footnote{Fig. 4 in~\cite{Chen2013endurance} shows that the relation between the endurance and the probability of stuck-at defects might be modeled by the lognormal distribution. However, only 25 cells were observed, which would not be enough to characterize the channel model.} Hence, the performance evaluation would need to be based on the experiments with real resistive memory chips.

%% Appendix:
%% If needed a single appendix is created by
%\appendix
%% If several appendices are needed, then the command
%\appendices
%% in combination with further \section-commands can be used.

%\appendix

%%% Use \section* for acknowledgement
%\section*{Acknowledgment}
%
%The authors would like to thank various sponsors for supporting
%their research.

%% References:
%% We recommend the usage of BibTeX:
%%

\IEEEtriggeratref{20}

\bibliographystyle{IEEEtran}
\bibliography{IEEEabrv,LWC_bib}

\end{document}